\documentclass[a4paper,preprintnumbers,showpacs,twocolumn,superscriptaddress,nofootinbib,amsmath,amssymb,floatfix]{revtex4-1}

\usepackage{amsmath,amssymb,graphicx,amsfonts}
\usepackage{mathrsfs}
\usepackage{comment}
\usepackage{xcolor}
\usepackage[shortlabels]{enumitem}
\usepackage{url}

\usepackage{graphicx}

\usepackage{dcolumn}
\usepackage{bm}
\usepackage{wasysym}
\usepackage{color}
\usepackage{multirow}
\usepackage{booktabs}
\usepackage{float}

\begin{document}
\def\Dt{\Delta t}
\def\Rdot{\stackrel{\cdot }{\relR}}
\def\Rddot{\stackrel{\cdot \cdot }{\relR}}


\title{\textbf{Small cross section of the synthesis of darmstadtium in the  $^{48}$Ca+$^{232}$Th reaction}}

\author{A.~K. Nasirov}
\email{Corresponding author: nasirov@jinr.ru}
\affiliation{Joint Institute for Nuclear Research, Dubna 141980, Russia}
\affiliation{Institute of Nuclear Physics, Tashkent 100214, Uzbekistan}
\author{A. Yusupov}
\email{yanuclear009@gmail.com}
\affiliation{Institute of Nuclear Physics, Tashkent 100214, Uzbekistan}
\author{B.~M. Kayumov}
\email{b.kayumov@newuu.uz} 
\affiliation{New Uzbekistan University, Tashkent 100000, Uzbekistan}
\affiliation{Institute of Nuclear Physics, Tashkent 100214, Uzbekistan}



\begin{abstract}
The smallness of the cross section of evaporation residues formed in the hot fusion reaction $^{48}$Ca+$^{232}$Th is analyzed by the dinuclear system model (DNS). The capture probability has been calculated by solving the dynamical equations of motion for the relative distance between the centers-of-mass of the DNS nuclei. Fusion of nuclei is considered as evolution of the DNS to a stable compound nucleus. The fusion probability has a bell-like shape and quasifission is one of reasons causing smallness of the yield of the evaporation residues products. Another reason is the decrease of the fission barrier for the isotopes $^{275-285}$Dm related with the shell effects in the neutron structure. The agreement of the theoretical results obtained for the yield of the evaporation residues with the experimental data measured in the Factory of superheavy elements of Joint Institute for Nuclear Research is well.
\end{abstract}
\maketitle
\section{Introduction}
The synthesis of new superheavy elements as a result of collisions of heavy nuclei is one of the important topics for many people interested in modern nuclear physics \cite{Hofman2000}. In recent years, new superheavy elements have been synthesized using heavy elements and actinide nuclei from the periodic table \cite{Armbruster2000,Morita2004, Oganessian2007}.

Currently, there is no theoretical model that fully explains the process of complete fusion in heavy ion collisions. Existing nuclear models are able to explain some features of this process. Several mechanisms are analyzed to explain the joining processes. To date, many theoretical studies have been carried out to calculate the evaporation residue (ER) cross sections in the complete fusion reaction at heavy ion collisions \cite{Antonenko1995,Adamian2000,Zagrebaev2008,Wang2012,Hong2017}.    The cross section of the evaporation residue depends on the collision energy and orbital angular momentum of the entrance channel and the physical properties of the projectile-target pair. A knowledge about the fusion mechanism  is very useful at the exploration of the optimal conditions for the synthesis of new superheavy elements. It is well known that the  ER cross section of the heaviest elements is very small and its excitation function range  is very narrow \cite{Armbruster2000,Oganessian2005}. Only in a narrow range 15--20 MeV of the collision energy values corresponds to the observable excitation functions of synthesis of the superheavy elements. To determine the best conditions for the input channel during the synthesis of superheavy elements, theoretical calculations usually study the dependence of the evaporation cross section  $\sigma_{ER}$ on the collision energy ($E_{\rm c.m.}$), orbital angular momentum ($L=\ell\hbar)$ and structure of colliding nuclei \cite{Antonenko1995,Giardina2000}.
\begin{eqnarray}
    \sigma_{ER}\left(E_{\rm c.m.}\right)&=&\sum_{\ell=0}^{\ell_d}(2\ell+1)
    \sigma_{\rm cap}(E_{\rm c.m.},\ell)\nonumber\\
    &\times&P_{\rm CN}(E_{\rm c.m.},\ell) W_{\rm sur}(E_{\rm c.m.},\ell),
    \label{ER1}
\end{eqnarray}
where $P_{\rm CN}(E_{\rm c.m.},\ell)$ is a hindrance factor to complete fusion \cite{Antonenko1995,Giardina2000};  $\sigma_{\rm cap}(E_{\rm c.m.},\ell)$ is partial capture cross section and $W_{\rm sur}(E_{\rm c.m.},\ell)$ is survival probability of the rotating and heated compound nucleus against fission by neutron emission.

It is important to estimate accurately the cross section of complete fusion leading to form a compound nucleus. 
\begin{equation}
    \sigma_{\rm fus}\left(E_{\rm c.m.}\right)=\sum_{\ell=0}^{\ell_d}(2\ell+1)
    \sigma_{\rm cap}(E_{\rm c.m.},\ell)P_{\rm CN}(E_{\rm c.m.},\ell).
    \label{CF}
\end{equation}

In heavy ion collisions  with massive nuclei the ER cross section is very small part of the fusion cross section. Its experimental value is determined as a sum of the cross sections of the measured yield of the reaction products of the fusion-fission and evaporation residue channels:
\begin{equation}
    \sigma^{(\rm exp)}_{\rm fus}\left(E_{\rm c.m.}\right)=
    \sigma^{(\rm exp)}_{\rm fus.fis}\left(E_{\rm c.m.}\right)+
    \sigma^{(\rm exp)}_{\rm ER}(E_{\rm c.m.}).
    \label{Cexp}
\end{equation}

It is well known that in some reactions the mass distribution of the  fusion-fission products may overlap with the one of the quasifission products. In this case, it is necessary to separate pure fusion-fission products from quasifission products. It is an ambiguous  task at the analysis of the measured data \cite{Giardina2018}. As a result the values of the fusion probability $P_{\rm CN}$ extracted from the measured data may be incorrect \cite{Loveland2015}. This creates uncertainty at estimation of $P_{\rm CN}$ basing in the yield of the binary products. One of reason causing smallness of the ER cross section is the dominance of the quasifission events in the reactions with the massive nuclei.
 
\begin{figure}[ht]
\centering
\includegraphics[width=0.9\linewidth]{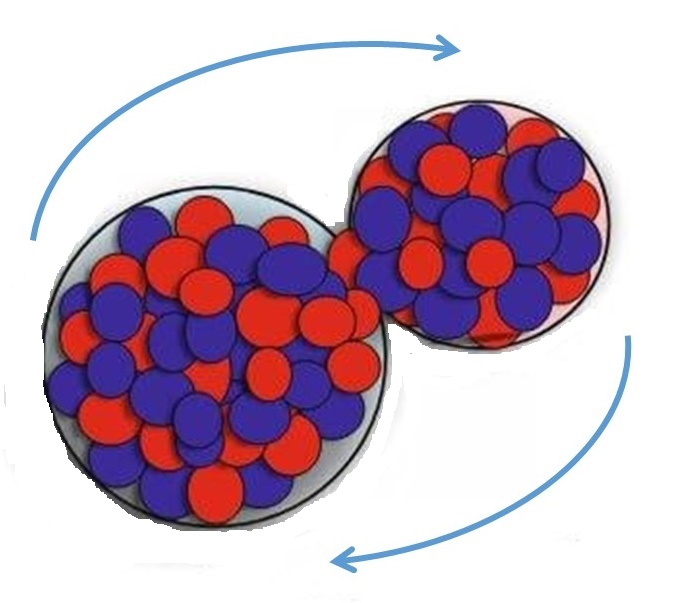}
\caption{(Color online) The sketch of the nucleon transfer between fragments of the rotating DNS 
formed at capture of the projectile nucleus by the target nucleus.}
\label{capDNS}
\end{figure}

In this work, we used the DNS model for the description of the ER cross sections. According to the DNS model,  a “neck” appears between the close surfaces of the projectile and the target nuclei and through this “neck” nucleons are transferred between nuclei (Fig. \ref{capDNS}). During the process of mutual transfer of nucleons, the nucleons occupy the empty quantum states of the acceptor-nucleus \cite{Hamilton2013}.

 The study of the fusion mechanism is also directly related with the PES of the system. It allows us to evaluate possible reaction channels and to estimate the excitation energy of the DNS. During the fusion process, nucleons move from the light nucleus to the heavy one. The motion of the nucleons takes place as a diffusion process, but the average flow goes in one  or another direction as a function of the PES landscape which determines by the DNS angular momentum. Nucleon transfer changes the mass and charge distribution in the DNS fragments and there is a possibility of the DNS breakup which is considered as quasifission process. Experimental and theoretical analysis of the yield of the reaction products  shows   that  the process of quasifission is the dominant channel compared to complete fusion \cite{Munzenberg2015}. 
  
The main goal of this research work is to calculate the cross section of the capture, complete fusion and evaporation residue formation for the theoretical study of the synthesis of superheavy element $^{280-xn}$Ds in the $^{48}$Ca+$^{232}$Th reaction. In the recent experiments of the Flerov Laboratory of Nuclear Reaction of JINR 
(Dubna, Russia), the maximum values of the cross sections 0.7$^{+1.1}_{-0.5}$ pb 
for the 4n channel \cite{Oganessian2023} and 0.34$^{+0.59}_{-0.16}$ for the 5n channel \cite{Oganessian2024}  of the evaporation residues at
the synthesis of the element Ds have been measured. These cross sections are 
much smaller than the experimental data 15$^{+9}_{-6}$ obtained  in the cold fusion $^{64}$Ni+$^{208}$Pb  reaction \cite{Hofmann1998}. Note these cold and hot fusion reactions lead to the different isotopes of Ds. It is important for researchers to know the reasons unambiguously leading to this difference in the observed ER cross sections. Is it related with the formation of the compound nucleus or/and its survival probability against fission.  

The process of synthesis of superheavy elements is considered as a final of the three stages. The first stage represents a competition between the deep-inelastic collision (this process occurs due to incomplete momentum transfer) and the capture process (this process occurs due to full momentum transfer). If the kinetic energy of the projectile is greater than the Coulomb barrier of the nucleus-nucleus interaction potential,  deep inelastic collision or capture projectile by the target nucleus occurs after dissipation of the sufficient part of the relative kinetic energy. If the DNS  is able to overcome the Coulomb barrier from the inside part the potential shell to outside then deep-inelastic collision takes place. Otherwise, the DNS nuclei is captured by the potential interaction well. The last condition is called full momentum transfer. The competition between capture and deep-inelastic collisions depends on the charge and mass numbers of the colliding nuclei, relative energy and orbital angular momentum  of collision.

At the second stage of the evolution of the DNS there is a competition between formation of a compound nucleus (complete fusion of nuclei) or the DNS breakup into two parts without reaching the equilibrium state of the compound nucleus. At the last stage, the heated and rotating compound nucleus should survive against fission process.

In this work involving the $^{48}$Ca$+^{232}$Th reaction, we calculated the cross sections for the capture of nuclei, fusion and evaporation residue of the resulting compound nucleus depending on various energies and orbital angular momentum.

\section{Capture cross section}

Calculation of the capture probability of the projectile by target nucleus at energies near the Coulomb barrier is performed by solution dynamical equations for the relative motion of the incoming trajectory of collision, as a result partial capture cross section can be calculated by using followed equation:
\begin{equation}
\sigma_{\rm cap}(E_{\rm c.m.}, \ell, \{\alpha_i\})=\dfrac{\lambda^2}{4\pi}\sum_{\ell=0}^{\ell_d}(2\ell+1)P_{\rm cap}^\ell(E_{\rm c.m.}, \ell, \{\alpha_i\})
    \label{4}
\end{equation}
where $\lambda$ is the de Broglie wavelength of the input channel and $P_{\rm cap}(E_{\rm c.m.}, \ell;\ \{\alpha_i\})$ is the capture probability; $\ell_d$ is the dynamical maximum value of the orbital momentum leading to capture in collisions with the orientation angles ${\alpha_i}$ of the axial symmetry axis of colliding nuclei relative the beam direction. The capture probability ${P}_{\ell}^{cap}(E)$ can be equal to 1 or 0 for the given beam energy and orbital angular momentum. For the given energy the number of partial waves ($\ell_d$) leading to capture is calculated  by solution of equations of the relative motion of nuclei \cite{Giardina2000,Nasirov2005}:
\begin{eqnarray}
 \label{maineq} &&\frac{d \dot R}{dt} +
 \gamma_{R}(R,\alpha_1,\alpha_2)\dot R(t)= F(R),\\
 \label{maineq2} &&F(R,\alpha_1,\alpha_2)=
 -\frac {\partial V(R,\alpha_1,\alpha_2)}{\partial R}-
 \dot R^2 \frac {\partial \mu(R)}{\partial R}\,,\\
 \label{maineq3}&&\frac{dL}{dt}=\left(\dot{\theta}
 R(t) -\dot{\theta_1} R_{1eff} -\dot{\theta_2} R_{2eff}\right)\nonumber \\
 &&\hspace{0.625 cm}\times\gamma_{\theta}(R,\alpha_1,\alpha_2)R(t),\\
 &&L_0=J_R(R,\alpha_1,\alpha_2) \dot{\theta}+J_1 \dot{\theta_1}+J_2 \dot{\theta_2}\,,\\
 &&E_{rot}=\frac{J_R(R,\alpha_1,\alpha_2) \dot{\theta_{}}{}^2}2+\frac{J_1
 \dot{\theta_1}^2}2+\frac{J_2 \dot{\theta_2}^2}2\,,
 \end{eqnarray}
 where $R\equiv R(t)$ is the relative distance between centres of mass interacting nuclei; $\dot R(t)$ is the corresponding velocity; $L_0$ ($L_0=\ell_0 \hbar$) and $E_{\rm kin}=E_{\rm c.m.}$ at $R \rightarrow{\infty}$ are initial conditions;  $\mu(R,\alpha_1,\alpha_2)=m A_P A_T/(A_P+A_T)$, where  $Z_P(A_P)$ and $Z_T(A_T)$ are charge (mass) numbers of the colliding nuclei, respectively, $m$ is mass of nucleon; $J_R=\mu R^2$  and $J_i=A_i m(a_i^2+b_i^2)/5$ are  moment of inertia of the DNS and its fragments, respectively;  $\dot\theta_i$ is the angular velocity of the fragment ``i'', $i=1,2$;  $\dot\theta$ is the angular velocity of the whole DNS around its centre of mass. $\gamma_{R}$ and  $\gamma_{\theta}$ are the friction coefficients for the relative motion along $R$ and the tangential motion when two nuclei roll on each other's surfaces, respectively. Their values are determined from the estimation of the particle-hole excitation in nuclei and nucleon exchange between them \cite{Giardina2000,Nasirov2005,Adamian1997}. $V(R,\alpha_1,\alpha_2)=V(Z_P,A_P,Z_T,A_T,R,\alpha_1,\alpha_2)$ is the  nucleus-nucleus potential calculated by the double folding procedure \cite{Giardina2000,Nasirov2005}  with the effective nucleon-nucleon forces suggested by Migdal \cite{Migdal1983}. 
\begin{figure}[ht]  
\centering
\includegraphics[width=1.0\linewidth]{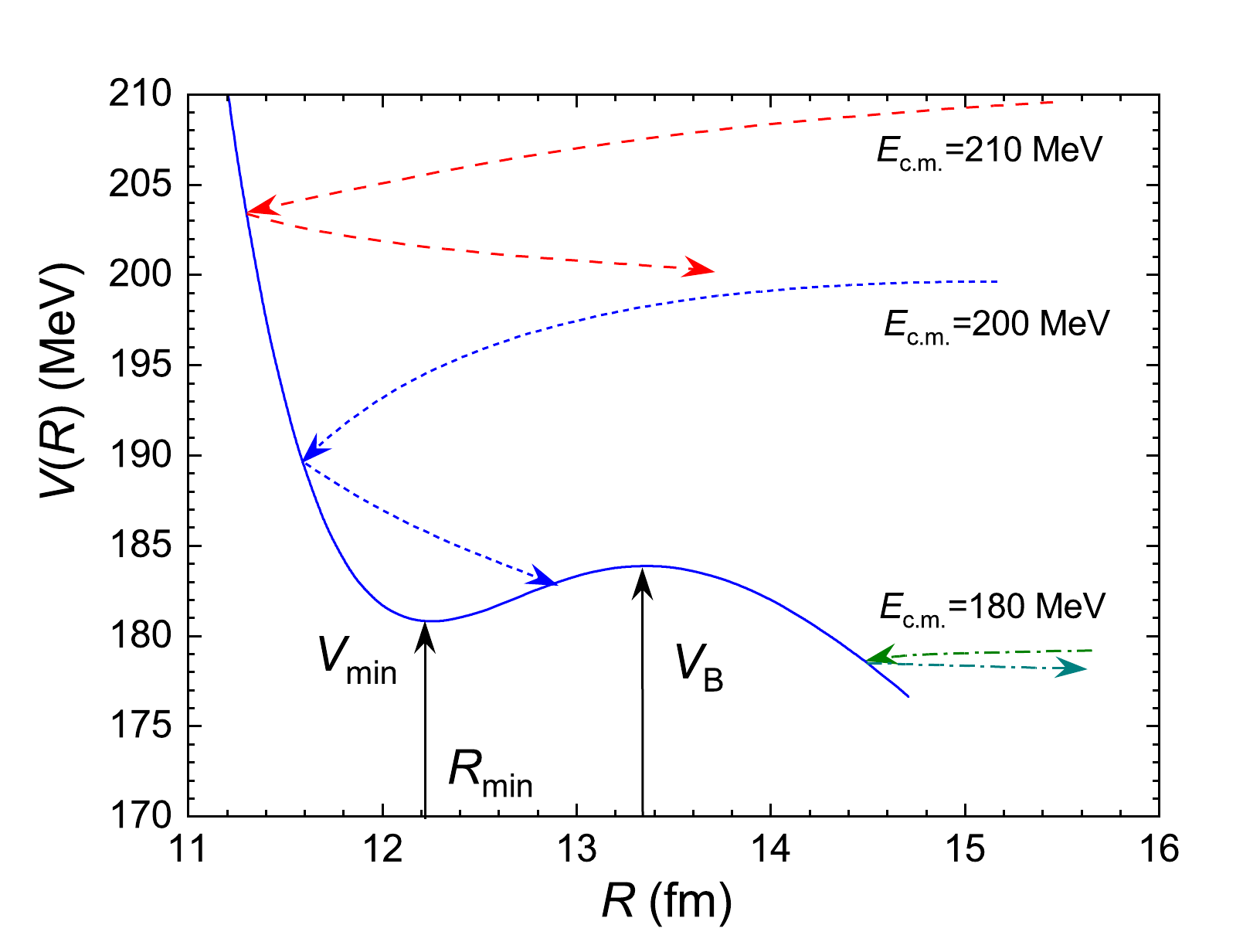}
\vspace{-0.25 cm}
\caption{(Color online) The sketch of trajectories of the inelastic collisions (at $E_{\rm c.m.}$=210 and 180 MeV) and capture (at 200 MeV) is due to a decrease in the initial kinetic energy of collision under the influence of friction forces. The nuclear interaction potential $V(R)$ is shown by the solid curve. }
\label{trajec}
\end{figure}

Figure \ref{trajec} shows the trajectories of deep-inelastic collisions at energies $E_{\rm c.m.}=210$ and 180 MeV, according to the DNS model, when the full momentum transfer of the relative motion does not take place. In collisions with the energies around $E_{\rm c.m.}=200$ MeV the full momentum transfer occurs, i.e. projectile nucleus is captured by the target nucleus. The kinetic energy is fully dissipated and the arrow of energy is ended in the bottom of the well of the nucleus-nucleus interaction $V_{\rm min}(R_{\rm min}, \ell,\{\alpha_i \})$ at the distance $R=R_{\rm min}$. 

The depth of the interaction potential $B_{\rm qf}=V_B-V_{\rm min}$ is used in the DNS model as quasifission potential. It determines the stability of the DNS against  to breakup into two parts (quasifission). Therefore, it is called a quasifission barrier (see Fig. \ref{trajec}). The size of the potential well depends on the mass and charge numbers of interacting nuclei, and on the orientation angles $\alpha_i$ of the symmetry axis of the deformed nucleus ($|\beta_i|>0$) relative to the direction of the projectile velocity ($i= 1$ (projectile), 2 (target)) \cite{Nasirov2005} and on the orbital angular momentum ($\ell$). 

For the given relative energy ($E_{\rm c.m.}$) and orbital angular momentum ($\ell$), the capture cross section ($\sigma_{\rm cap}$) can be expressed as the sum of the competing channels cross sections:
\begin{eqnarray}
    \sigma_{\rm cap}(E_{\rm c.m.}, \{\alpha_i\})&=&\sigma_{\rm fus}(E_{\rm c.m.}, \ell, \{\alpha_i\})\cr
   & +&\sigma_{\rm qfis}(E_{\rm c.m.}, \ell, \{\alpha_i\})\cr
   &+&\sigma_{\rm ffis}(E_{\rm c.m.}, \ell, \{\alpha_i\}),
    \label{3}
\end{eqnarray}
where $\sigma_{\rm fus}$ is the cross section for complete fusion of nuclei; 
$\sigma_{\rm qfis}$ is the cross section for DNS quasifission, $\sigma_{ffis}$ is 
the cross section for fast fission of the rotating mononuclear.

The potential well gives the possibility of capture. If the value of the orbital momentum and energy satisfy following conditions: $\ell<\ell_d$ and $E_{\rm c.m.}>V_B$, then the capture probability $P_{\rm cap}$ is equal to unity. Conversely, if $\ell>\ell_d$ and $E_{\rm c.m.}>V_B$, then $P_{\rm cap}$ is equal to zero. 

However, in the case $E_{\rm c.m.}<V_B(\ell)$ the capture process can occurs due to the tunneling effect where the value of $E_{\rm c.m.}$ satisfies the condition $V_{\rm min}(\ell)<E_{\rm c.m.}<V_B(\ell)$ for any value of the $\ell$-orbital angular momentum. Then the probability of the barrier penetrability $\mathcal{P}^{(\ell)}_{\rm tun}$ is calculated by the WKB formula obtained in Ref. \cite{Kemble1935}:
\begin{eqnarray}\label{penetration}
\mathcal{P}^{(\ell)}_{tun}(E_{\rm c.m.},\{\alpha_i\})=\frac{1}
{1+\exp\left[2K(E_{\rm c.m.},\ell,\{\alpha_i\}\})\right]},
\end{eqnarray}
where
\begin{eqnarray}
K(E_{\rm c.m.},\ell,\{\alpha_i\})&=&\int\limits_{R_{in}}^{R_{out}} dR\nonumber\\
&\times&\sqrt{\frac{2\mu}{\hbar^2}(V(R,\ell,\{\alpha_i\})-E_{\rm c.m.})},\nonumber\\
\end{eqnarray}
$R_{\rm in}$ and $R_{\rm out}$ are inner and outer turning points which were estimated by $V(R)=E_{\rm c.m.}$.

\section{Potential energy surface and driving potential}

\begin{figure}[ht]
\hspace{-0.85 cm}
\includegraphics[width=1.0\linewidth]{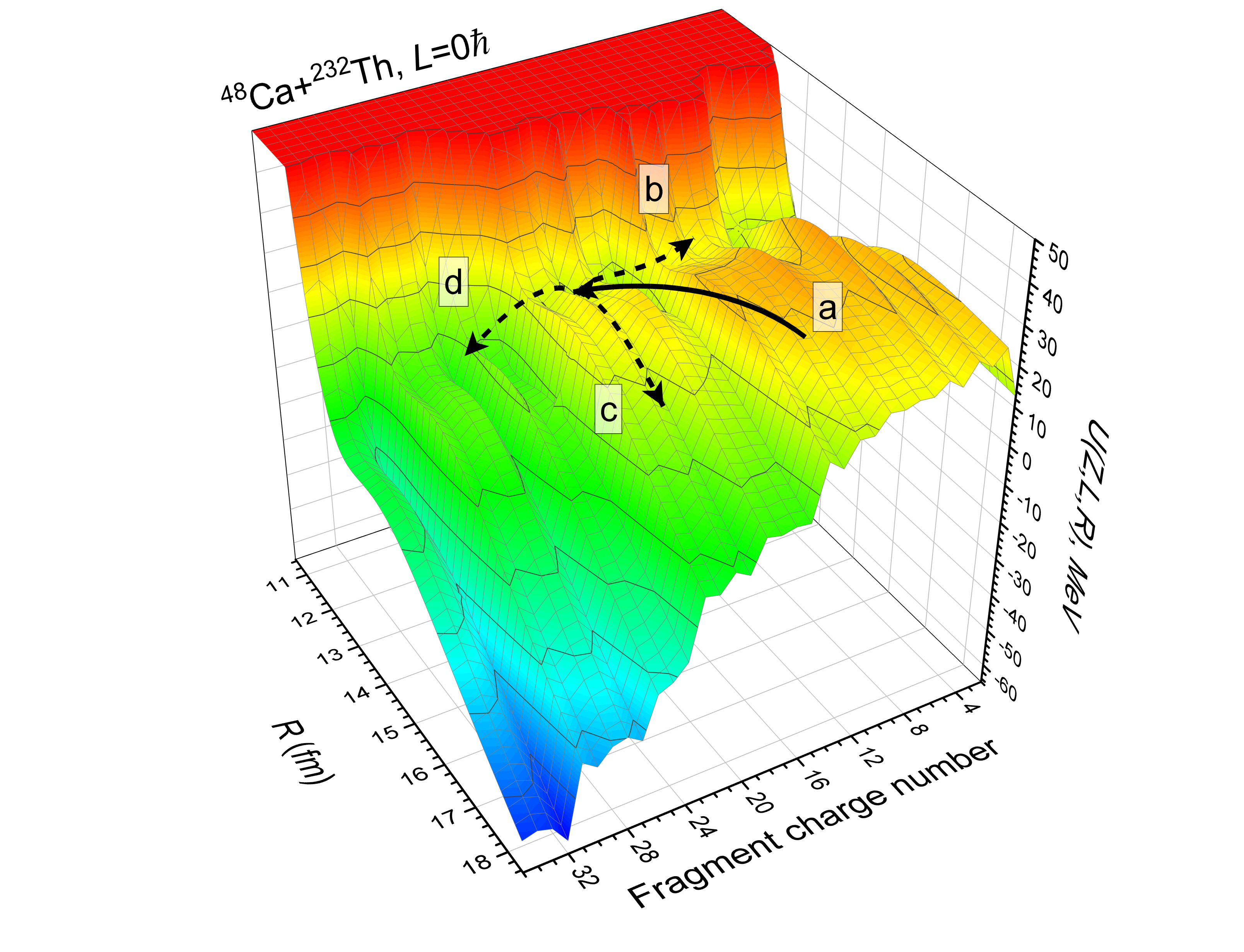}
\caption{The potential energy surface  of the DNS formed in the reaction $^{48}\text{Ca}+^{232}\text{Th}$ calculated for collisions with the value of the orbital angular momentum $\ell=0$ and  the orientation angles $\alpha_1= 30^\circ$ and $\alpha_2=135^\circ$. The arrow (a) shows the input capture channel; the arrow (b) shows  the directions of the complete fusion by the nucleon transfer from a light nucleus to a heavy one; arrows (c) and (d) show the directions of the DNS decay into mass-asymmetric and symmetric quasifission channels, respectively.}
\label{PES}
\end{figure}

The main role in the formation and evolution of the DNS is played by the potential energy surface (PES) and dynamic coefficients (friction force and moment of inertia) \cite{Brewer2018}. PES is a function of the mass $(A_i)$ and charge $(Z_i)$ numbers of the colliding nuclei, the orbital angular momentum ($\ell$) and the relative distance between their centers of mass, and can be calculated as a function of $R$ (See Figure \ref{PES}):
\begin{eqnarray}
    U(Z, A, \ell, R, \{\alpha_i\}) &=& V(Z, A, \ell, R, \{\alpha_i\})+Q_{gg} \cr
    &-&V_{rot}^{CN},
    \label{6}
\end{eqnarray}
where $Q_{gg}(Z,A) = B_1(Z,A) + B_2(Z_c,A_c) - B_{\rm CN}$ is the reaction balance energy; $B_1$, $B_2$ and $B_{\rm CN}$ are binding energy of the interacting nuclei ($Z, A; Z_c, A_c$) of DNS and the compound nucleus with the charge $Z_{\rm CN}$ and mass $A_{\rm CN}$ numbers which is formed at complete fusion; their values are taken from the table in Refs. \cite{AUDI2003,Moller1995}; $Z_{\rm CN}=Z_P+Z_T, A_{\rm CN}=A_P+A_T$, $Z=Z_{\rm CN}-Z_c$ and $A=A_{\rm CN}-A_c$, where $Z$ and $A$ are charge and mass numbers of the light fragment of DNS and $Z_c$ and $A_c$ are the ones of the conjugate heavy fragment. The nucleus-nucleus interaction ($V(R)$) between the projectile and the target nuclei consists of three parts:
\begin{eqnarray}
   V(Z, A; R, \{\alpha_i\}) &=& V_N(Z, A; R, \{\alpha_i\})\cr
   &+&V_C(Z, A; R, \{\alpha_i\})\cr
   &+&V_{rot}(Z, A; R, \{\alpha_i\})
    \label{7}
\end{eqnarray}
where $V_N$, $V_C$ and $V_{rot}$ are the nucleus-nucleus, Coulomb and rotational potentials, respectively. Method of calculation of the potentials $V_N$ and $V_C$ have been  presented in Appendix A of Ref. \cite{Nasirov2005}, and the rotational part the nucleus-nucleus interaction $V_{rot}$ is given by: 
\begin{equation}
V_{\rm rot}(Z, A; R, \{\alpha_i\})=\frac{\ell_0(\ell_0+1)\hbar^2}{2\mu R^2+J_1+J_2}.
\end{equation}
The potential energy surface $V(Z, A; R, \{\alpha_i\})$ with the showed directions of the possible evolution of the DNS formed in the reaction $^{48}\text{Ca}+^{232}\text{Th}$ is presented in Fig. \ref{PES} for the collision with $L=0$, orientation angles $\alpha_1=30^o$ and $\alpha_2=135^o$. The arrow (a) shows the input capture channel; the arrow (b) shows  the directions of the complete fusion by the nucleon transfer from a light nucleus to a heavy one; arrows (c) and (d) show the directions of the DNS decay into mass-asymmetric and symmetric quasifission channels, respectively. When DNS breakup into two fragments before (without) reaching an equilibrium state of compound nucleus after capture is called quasifission. 

\begin{figure}[ht]
\centering
\includegraphics[width=0.98\linewidth]{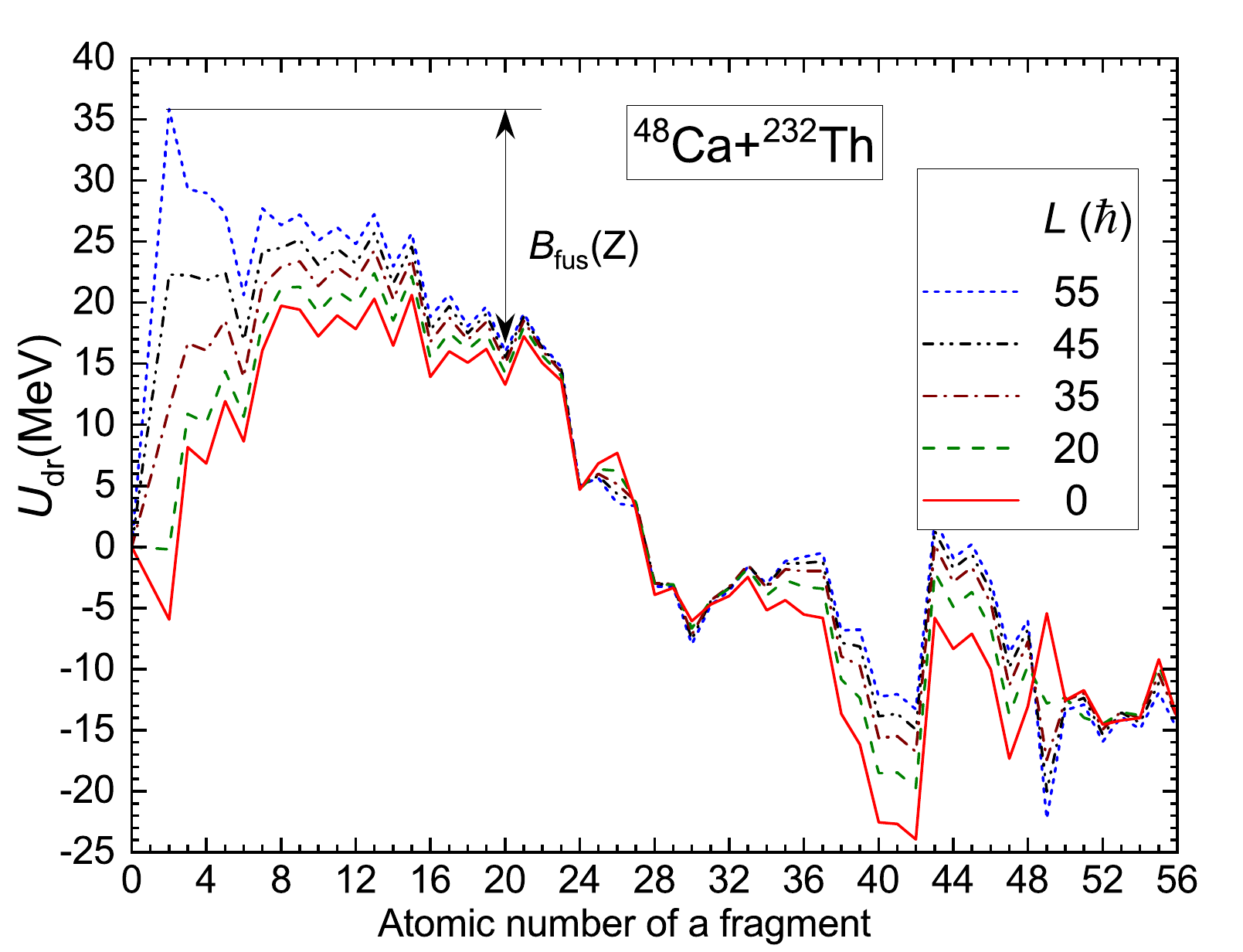}
\vspace{-0.25 cm}
\caption{(Color online) The driving potential calculated for the 
 $^{48}\text{Ca}+^{232}\text{Th}$ reaction for the orientation angles $\alpha_P=30^\circ$ and $\alpha_T=45^\circ$ of the axial symmetry of the nuclei of DNS formed with the angular momentum $L=0, 20, 35, 45, 55 \hbar$. $B_{\rm fus}^*(Z)$ is an internal nuclear barrier that causes hindrance to complete fusion for the charge asymmetry state $Z=20$ of the DNS.}
\label{UdrL}
\end{figure}

The curve connecting of the minima of the PES valley is used as driving potential $U_{\rm dr}$ for the DNS formed in the given reaction:
\begin{eqnarray}
    U_{\rm dr}(Z, A, \ell, \{\alpha_i\}) &=& V(Z, A, \ell, R_{\rm min}, 
    \{\alpha_i\})+Q_{gg} \cr
    &-&V_{rot}^{CN}.
    \label{Udr}
\end{eqnarray}
In DNS model, the intrinsic fusion barrier $B^*_{\rm fus}$ for the fusion from the charge asymmetry $Z$ of the DNS can be found from the peculiarities of $U_{\rm dr}$ as it is showed in Fig. \ref{UdrL} for the DNS with the angular momentum $L=55 \hbar$. Its values may vary depending on the deformation parameters $\{\beta_i\}$ of interacting nuclei and orientation angles $\{\alpha_i\}$.

\section{Fusion cross sections}

The extraction of the fusion probability  $P_{\rm CN}$ from the experimental data is ambiguous procedure since there is a freedom in definition of the capture $\sigma_{\rm cap}$ and fusion $\sigma_{\rm fus}$ cross sections \cite{Giardina2018,Loveland2015}:
\begin{equation}
  P_{\rm CN}(E_{\rm c.m.})=\sigma_{\rm fus}/(\sigma_{\rm cap}+\sigma_{\rm fus}).  
\end{equation}
This fact is related with the borders of the mass-energetic distributions 
of the capture events considered by the authors and possibility of the overlap of the mass distributions of the fusion-fission and quasifission products.  

The theoretical fusion probability  $P_{\rm CN}$ is calculated by the use of the intrinsic fusion barrier $B^*_{\rm fus}$, which hinders complete fusion and  quasifission barrier $B_{qf}$, which prevents the decay of DNS into two fragments. These barriers  are determined from the landscape of the potential energy surface and the estimation ways of them are shown in Figs. \ref{PES}(b) and \ref{PES}(c).

The fusion probability $P_{\rm CN}(E_{\rm c.m.},\ell,\{\alpha_i\})$ depends on the change in mass and charge distributions of $D_Z$ in DNS fragments after capture \cite{Kayumov2022}. In    general, it is calculated as the sum of the competing channel of quasifission and complete fusion at different charge asymmetries from the symmetric configuration $Z_{\rm sym}$ direction (d) in Fig. \ref{PES} of the DNS to the configuration corresponding to the maximum value of the driving potential $Z_{\rm max}$ and can be represented in this form:

\begin{eqnarray}
P_{\rm CN}(E_{\rm c.m.},\ell,\{\alpha_i\})
&=&\sum\limits^{Z_{max}}_{Z_{sym}}D_Z(E^*_Z,\ell,\{\alpha_i\})\nonumber\\
&\times&
P^{(Z)}_{\rm CN}(E^*_Z,\ell,\{\alpha_i\}).
\label{PcnDZ}
\end{eqnarray}
The values of the $D_Z(E^*_Z,\ell,\{\alpha_i\})$ are calculated by the solution of the transport master equation with the nucleon transition coefficients depending on the occupation numbers and energies of the single-particle states of nucleons of the DNS nuclei \cite{Kayumov2022}. The fusion probability $P^{(Z)}_{CN}(E^*_Z,\ell,\{\alpha_i\})$ for the DNS fragments with the charge configuration $Z$ rotating with the orbital angular momentum $\ell$ is calculated as the branching ratio of the level densities of the quasifission barrier $B^Z_{qf}(\ell,\{\alpha_i\})$ at a given mass asymmetry, over the intrinsic barrier $B^{*(Z)}_{fus}(\ell,\{\alpha_i\})$ and symmetry barrier $B^{(Z)}_{\rm sym}(\ell,\{\alpha_i\})$ on mass asymmetry axis  \cite{Nasirov2019}:
\begin{equation}
 \label{Pcnro} P^{(Z)}_{CN}(\xi)=\frac{\rho_{\rm fus}(\xi)}{\rho_{\rm fus}(\xi) +
\rho_{\rm qfiss}(\xi)+\rho_{\rm sym}(\xi)},
 \end{equation}
 where $\xi \equiv (E^*_Z,\ell,\{\alpha_i\})$ has been used for simplicity. The use of the level density function of the Fermi system leads to the formula for the fusion probability at the DNS excitation energy $E^*_Z$ and angular momentum $L$ from its charge asymmetry $Z$:
\begin{equation}
 \label{Pcn} P^{(Z)}_{CN}(\xi)=\frac{e^{-B_{\rm fus}^{*(Z)}/T_Z}}{e^{-B_{\rm fus}^{*(Z)}/T_Z} +
e^{-B_{\rm qfiss}^{*(Z)}/T_Z}+e^{-B_{\rm sym}^{*(Z)}/T_Z}}.
 \end{equation}
Here the values of the level density on the barriers  $B^{(Z)*}_{\rm fus}(\alpha_i)$, $B^{*(Z)}_{\rm sym}(\alpha_i)$ and $B^{(Z)}_{qf}(\alpha_i)$ have been used. To simplify the presentation of Eqs. (\ref{Pcnro})  and (\ref{Pcn}) the arguments $(\alpha_i)$ of the functions $E^*_Z(\alpha_i)$, $T_Z(\alpha_i)$, $B^{*(Z)}_{\rm fus}(\alpha_i)$, $B^{*(Z)}_{\rm sym}(\alpha_i)$ and $B^{(Z)}_{qf}(\alpha_i)$ are not indicated on their right sides of Eqs. (\ref{Pcnro})  and (\ref{Pcn}). $T_Z$ is the effective temperature of the DNS with the charge number $Z$ of its light fragment:
\begin{equation}
   T_Z= \sqrt{\frac{E^*_Z}{a}},
\end{equation}
where $a=A_{\rm CN}/12$ MeV$^{-1}$.  The excitation energy $E^*_Z(E_{\rm c.m},\ell)$ of the DNS with the charge $Z$ and mass $A$ numbers of the light fragment is determined by the difference between collision energy $E_{\rm c.m}$ and peculiarities of the driving potential $U_{\rm dr}$ calculated for the given value of $\ell$:
\begin{eqnarray}
E^*_Z(E_{\rm c.m},\ell,\alpha_i)&=&E_{\rm c.m}-V_{\rm min}(Z,A,R_m,\alpha_i)\nonumber\\
&-&\Delta Q_{gg}(Z,A),
\label{ExiZ}
\end{eqnarray}
where
\begin{eqnarray}
\Delta Q_{gg}(Z,A)&=&B_P(Z_P,A_P)+B_T(Z_T,A_T) \nonumber\\
&-&(B_1(Z,A)+B_2(Z_c,A_c)) 
\end{eqnarray}
is a change of the binding energy of the DNS fragments during its evolution from the initial value ($Z=Z_P$ and $A=A_P$) to the final configuration with the  charge and mass numbers $Z$ and $A$, respectively.

\begin{figure}[ht]
\centering
\includegraphics[width=1.0\linewidth]{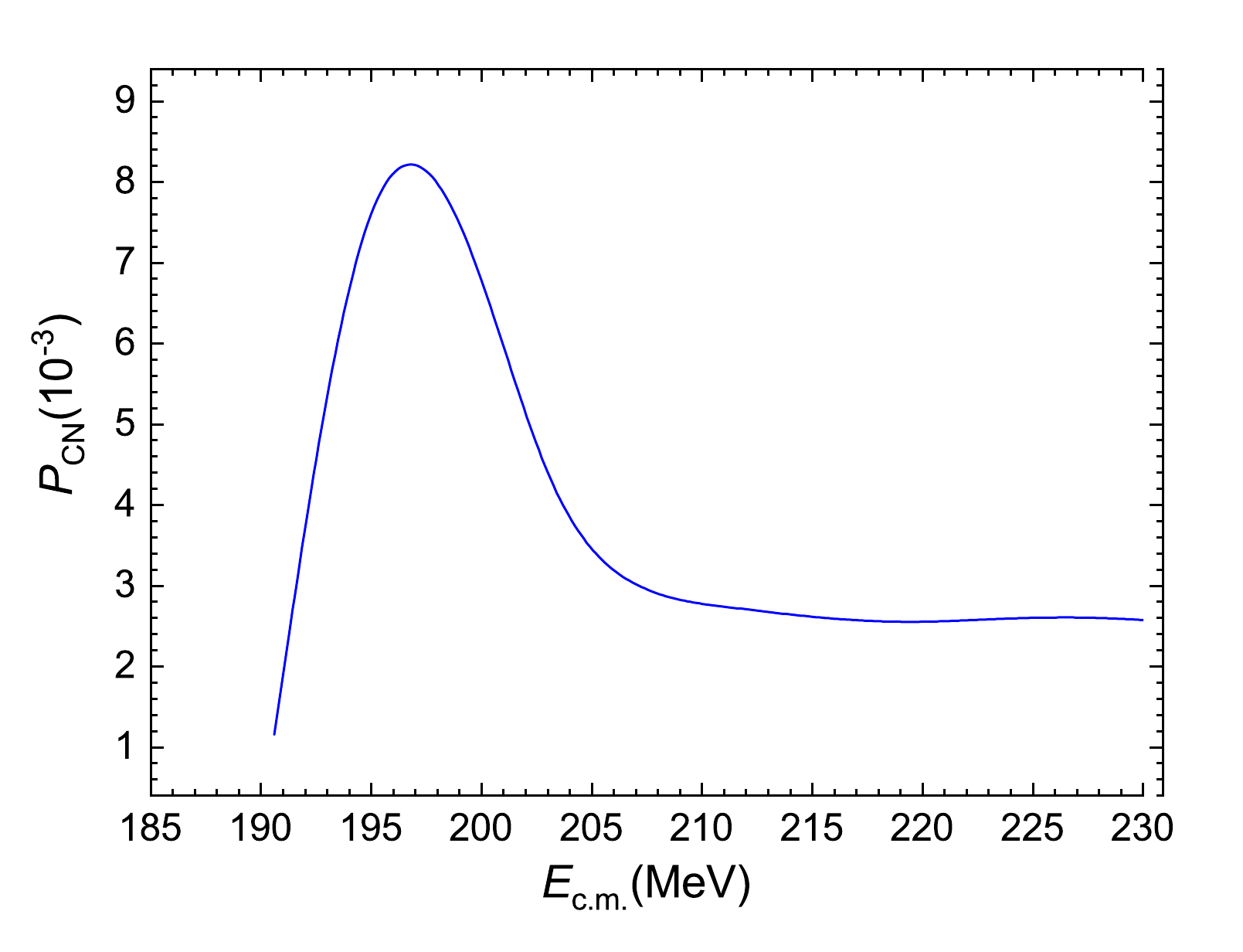}
\vspace{-0.75 cm}
\caption{(Color online) Dependence of the probability of complete fusion ($P_{\rm CN}$) on the collision energy in center of mass system for the reaction $^{48}\text{Ca}+^{232}\text{Th}$.}
\label{PcnF}
\end{figure}

\begin{figure}[ht]
\centering
\includegraphics[width=1.0\linewidth]{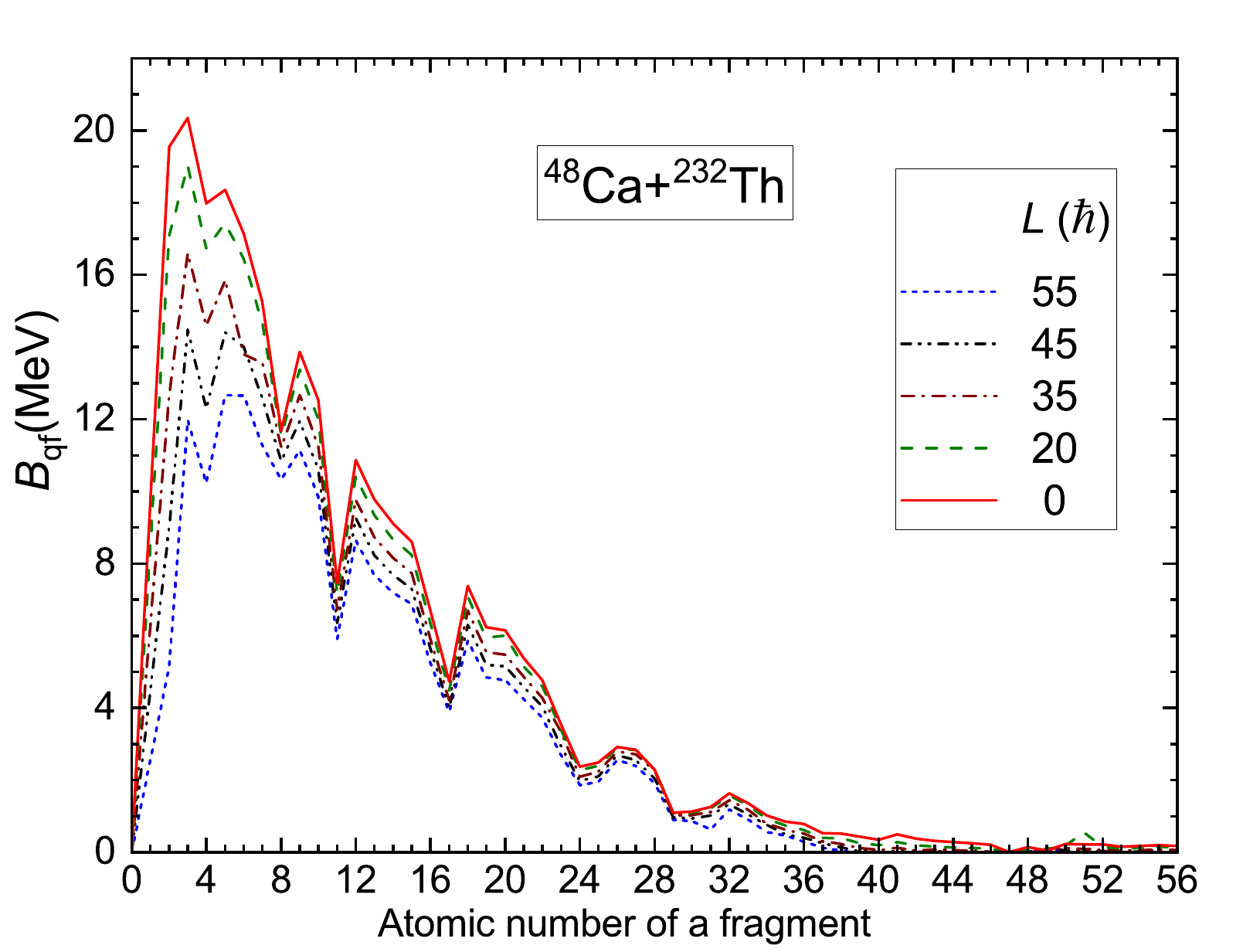}
\vspace{-0.35 cm}
\caption{(Color online) Dependence of the DNS quasifission barrier $B_{qf}$  
on the charge of the light fragment 
in the reaction $^{48}\text{Ca}+^{232}\text{Th}$.}
\label{BqfZ}
\end{figure}

The fusion cross section is calculated by Eq. (\ref{CF}) and its some part corresponding to the range $\ell> \ell_B$ contributes to the fast fission cross section:
\begin{equation}
    \sigma_{ffis}=\sum_{\ell=\ell_B}^{\ell_d}(2\ell+1)\sigma_{\rm fus}(E_{\rm c.m.}, \ell), 
    \label{ffis}
\end{equation}
where $\ell_d$ - the maximum value of the angular momentum leading to capture (DNS formation) at a given collision energy; $\ell_B$ is a value of $\ell$ at which the fission barrier of the compound nucleus disappears \cite{Sierk1986}. The fast fission is the fission of the rotating mononucleus  into two fragments due to disappearance of its fission barrier $B_f$ which depends on the angular momentum $L$ and excitation energy $E^*_{\rm CN}$ \cite{Kayumov2022}. The fast fission process does not allow the heated and rotating mononucleus to turn into a compound nucleus.

The results of averaging over all orientation angles $\alpha_T$  of the axis of axial symmetry of the target-nucleus $^{232}$Th in the interval $0^\circ\leq\alpha_T\leq90^\circ$:
\begin{eqnarray}
    \sigma_{\rm fus}(E_{\rm c.m.}, \ell)&=&\int\limits_0^{\pi/2}\sigma_{\rm cap}(E_{\rm c.m.}, \ell, \alpha_T) P_{CN}(E_{\rm c.m.}, \ell, \alpha_T) \cr
    &\times&\sin\alpha_T d\alpha_T
    \label{12}
\end{eqnarray}
are used in calculations of the partial cross sections of the evaporation residue formation. Surface vibrations relative to the cores in the ground spherical state are taken into account if one of colliding nucleus ($^{48}$Ca) has spherical shape in its ground state. The procedure of the averaging  over vibration states of the spherical shape has considered in \cite{Nasirov2024}.  

The capture cross section is averaged over all orientation angles $\alpha_T$  by the similar way:
\begin{eqnarray}
    \sigma_{\rm cap}(E_{\rm c.m.}, \ell)&=&\int\limits_0^{\pi/2}\sigma_{\rm cap}(E_{\rm c.m.}, \ell, \alpha_T)  \cr
    &\times&\sin\alpha_T d\alpha_T.
    \label{capalpha}
\end{eqnarray}

The partial cross section of complete fusion in the collision with the orientation angles $\alpha_i$ $(i=1,2)$  is calculated taking into account the hindrance to complete fusion caused by quasifission:
\begin{eqnarray}
    \sigma_{\rm fus}(E_{\rm c.m.}, \ell, \{\alpha_i\})&=&
    \sigma_{\rm cap}(E_{\rm c.m.}, \ell, \{\alpha_i\})\cr
    &\times& P_{CN}(E_{\rm c.m.}, \ell, \{\alpha_i\}).
    \label{fusl}
\end{eqnarray}

The theoretical values of the fusion probability to compare with the experimental data are found as a ratio of the total fusion and capture cross sections:
\begin{equation}
    P_{\rm CN}(E_{\rm c.m.})=\frac{\sigma_{\rm fus}(E_{\rm c.m.})}{\sigma_{\rm cap}(E_{\rm c.m.})},
\end{equation}  
where the total cross sections $\sigma_i$ ($i=$ fus, cap)  are calculated by summing the contributions of the partial waves leading to capture: $\ell$=0---$\ell_d$. The results of calculation  $P_{\rm CN}$ are presented in Fig. \ref{PcnF}. Its maximum values are in the range $E_{\rm c.m}=$192--205 MeV which is significantly higher than the Coulomb barrier $V_B=184$ MeV which is showed in Fig. \ref{trajec}. This result is calculated for collisions with the orientation angle $\alpha_T=30^o$. The value of $V_B$ is 198 MeV for the collisions with the orientation angle $\alpha_T=90^o$. The difference between $V_B$ calculated for small values of $\alpha_T$ and $E_{\rm c.m.})$=197 MeV corresponding to the maximum value of $P_{\rm CN}$ is an indication to the presence of the intrinsic barrier $B^*_{\rm fus}$. The DNS  should overcome this barrier during its evolution to be transformed into a compound nucleus. This means that $B^*_{\rm fus}$ causes hindrance to complete fusion and it increases events of the quasifission as the breakup of the DNS into two fragments due to small values of the quasifission barrier presented in Fig. \ref{BqfZ}.

The quasifission cross sections is calculated from the capture cross section by the expression:
\begin{equation}
    \sigma_{\rm qf}\left(E_{\rm c.m.}\right)=\sum_{\ell=0}^{\ell_d}(2\ell+1)
    \sigma_{\rm cap}(E_{\rm c.m.},\ell)(1-P_{\rm CN}(E_{\rm c.m.},\ell)).
    \label{QF}
\end{equation}

\begin{figure}[ht]
\centering
\includegraphics[width=1.\linewidth]{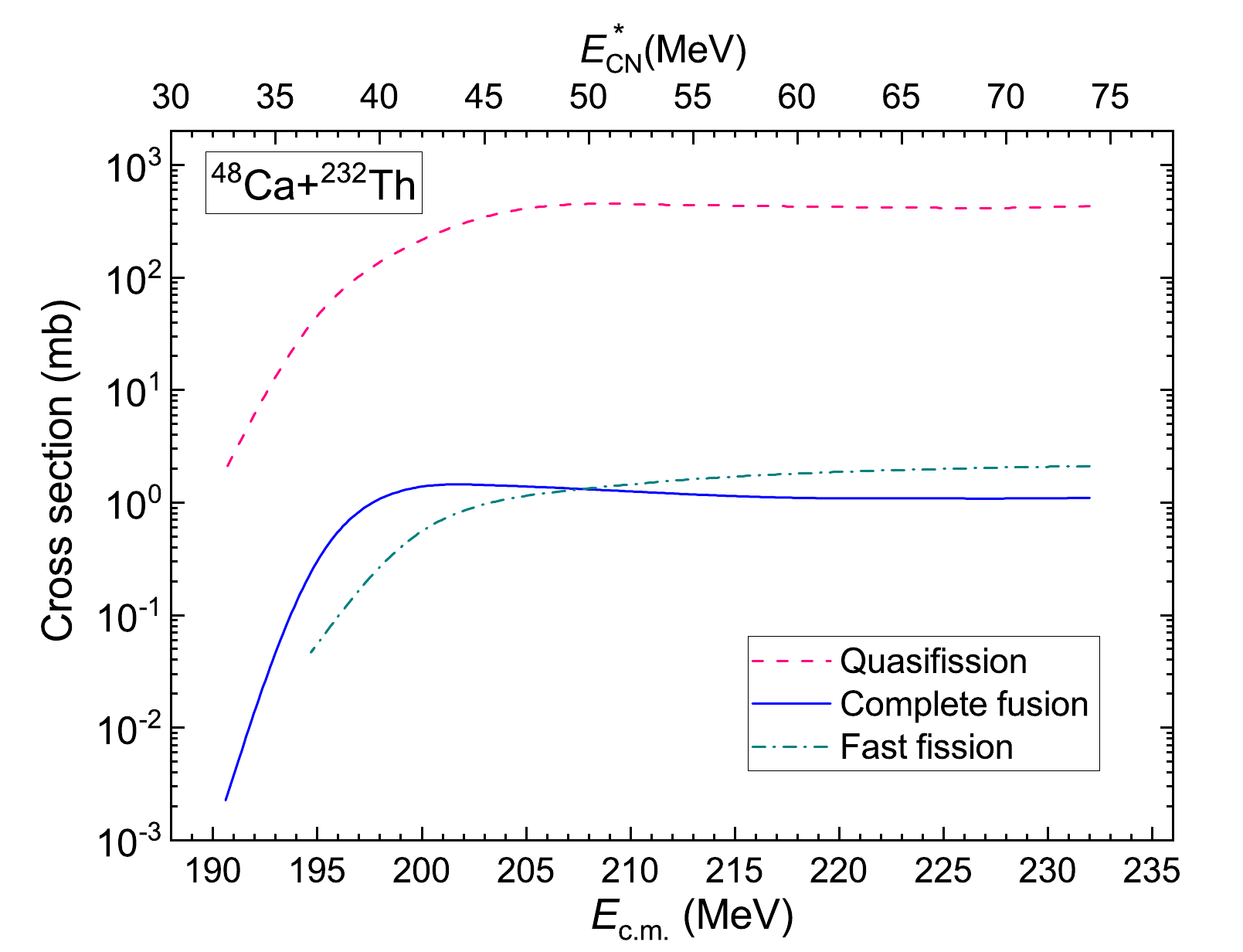}
\vspace{-0.5 cm}
\caption{(Color online) Cross sections of the quasifission (dashed curve), complete fusion (solid curve) and fast fission (dot-dashed curve) calculated in this work for the reaction $^{48}\text{Ca}+^{232}\text{Th}$. }
\label{CrossSection}
\end{figure}
The results of calculated cross sections of the complete fusion by Eq. (\ref{CF}), quasifission by Eq. (\ref{QF}), and fast fission by Eq. (\ref{ffis}) for the reaction $^{48}\text{Ca}+^{232}\text{Th}$ are presented in Fig. \ref{CrossSection}.
As it is seen from this figure,  the quasifission cross section is  dominant process and the capture cross section is determined by the quasifission cross section.  The fusion cross section increases from small values up to large values in the optimal range is $E_{\rm lab}=230-260$  MeV. Then fast fission becomes dominant due to the fact that the inequality $\ell_f<\ell$ begins to hold. The fission barrier of the heated and rotating heavy nucleus decreases  by the increase of its orbital angular momentum. At a certain value of the orbital angular momentum, the fission barrier completely disappears, the mononucleus loses stability and disintegrates into two fragments \cite{Sierk1986}. The fission barrier of heavy elements $(Z>106)$   exists only due to the quantum shell effects of the nuclear structure.   This kind of barrier decreases by the increase of the CN excitation energy   \cite{Kim2015,Mandaglio2018}.
The difference between fast fission and quasifission is that, a mononucleus undergoing rapid fission in a system that has survived against quasifission. According to the results obtained by the DNS model, the quasifission can occur for all values of its angular momentum $L_{DNS}=\ell_{DNS}\times h$. This is one of the main differences between fast fission and quasifission \cite{Graeger2010}.
\begin{figure}[ht]
\centering
\includegraphics[width=1.05 \linewidth]{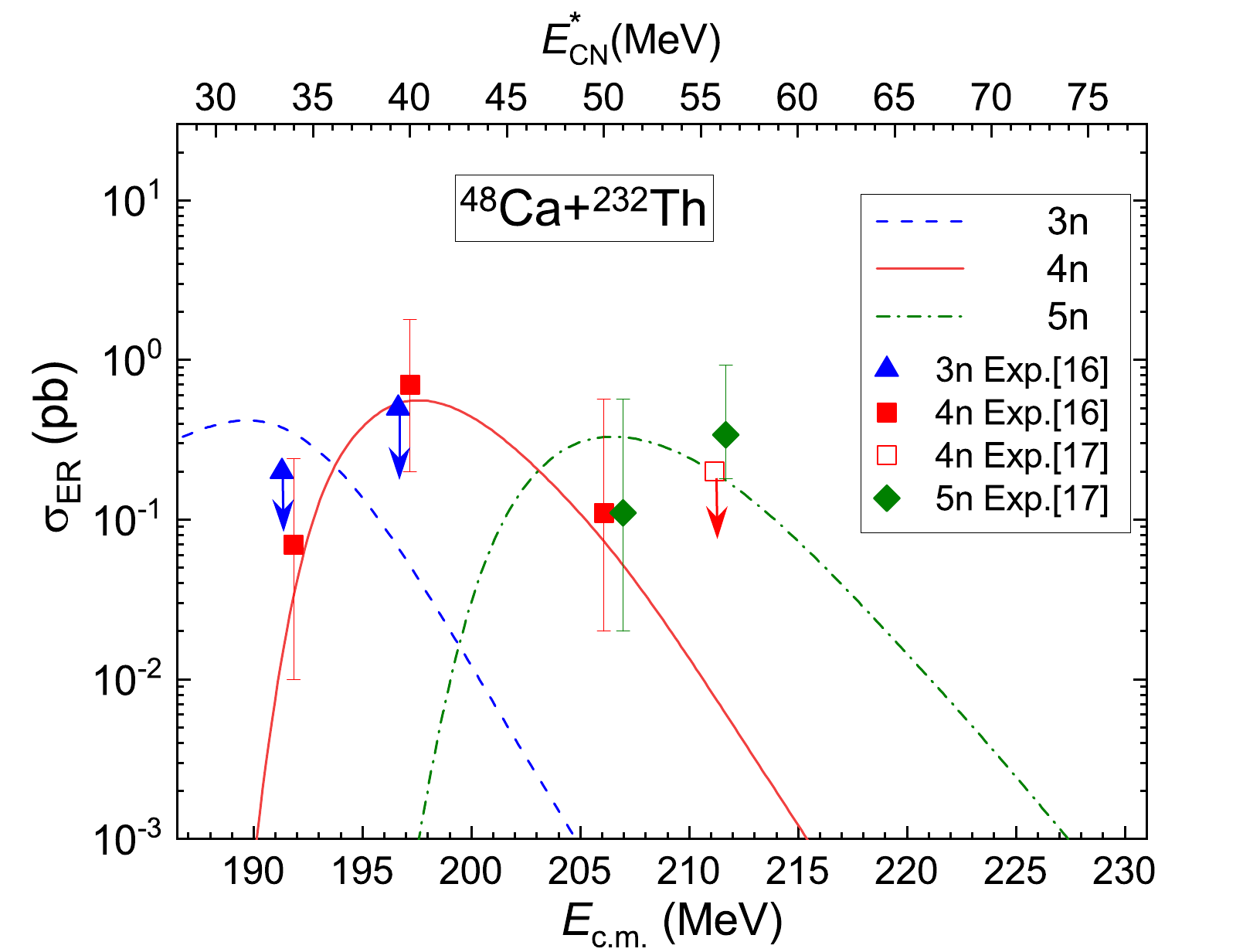}
\caption{(Color online) Comparison of the ER cross sections of the de-excitation 3n (dashed curve), 4n (solid curve) and 5n (dot-dashed curve) channels calculated in this work with the experimental data of the 3n (triangles) and 4n (solid squares) channels obtained in Ref. \cite{Oganessian2023} 
and of the 4n (open square) and 5n (diamonds) obtained in Ref. \cite{Oganessian2024}  of the de-excitation of the heated and rotating CN formed in the $^{48}\text{Ca}+^{232}\text{Th}$ reaction. The symbols with arrows show the
upper cross-section limits.}
\label{ERCrosSec}
\end{figure}

\section{Description of the experimental results of synthesis of Ds in hot fusion reaction}

The evaporation residue cross sections of the 3n, 4n and 5n de-excitation channels have been calculated by Eq. (\ref{ER1}). The  survival probability $W_{\rm sur}$ of the heated and rotating  CN is calculated  by the statistical model implanted in KEWPIE2 \cite{Kewpie2}, which is dedicated to the study of the evaporation residues at the synthesis of SHE (See Ref. \cite{Kayumov2022} for details of calculation of  $W_{\rm sur}$). The maximum of the evaporation residue cross section was observed for the 4n channel at $E_{\rm lab}$=237.5 MeV which corresponds to $E_{\rm CN}^*$=40 MeV of the CN excitation energy. Comparison of the theoretical results of this work 
with the measured data in Ref. \cite{Oganessian2023} shows that the description 
of the experimental data is well (see Fig. \ref{ERCrosSec}). 
 \begin{table}[ht]
\caption{The measured ER cross sections, excitation energy $E^*_{\rm CN}$ corresponding to the measured data and fission barrier $B_f$ of the CN calculated in Ref. \cite{Kowal2010}, as well as fusion probability $P_{\rm CN}$ for the cold fusion $^{64}$Ni+$^{208}$Pb (from \cite{Giardina2000}) and hot fusion $^{48}$Ca+$^{232}$Th  (this work) reactions.}
\begin{ruledtabular}
\begin{tabular}{ccccc}
Reaction & $\sigma_{\rm ER}$ (pb)  & $E^*_{\rm CN}$ (MeV)& $B_f$ MeV & $P_{\rm CN}$  \\
\hline
$^{64}$Ni+$^{208}$Pb & $15^{+9}_{-6}$          & 12.7 & 5.62 &  $10^{-5}$ \\
$^{48}$Ca+$^{232}$Th & $0.7^{+1.1}_{-0.5}$ & 40.37  & 3.29 & $8.3\times 10^{-3}$\\
\end{tabular}
\end{ruledtabular}
\label{ComHotCold}
\end{table}
It is interesting to establish reasons causing so strong difference (20 times) between the  measured cross sections of the isotopes of Ds in the cold fusion $^{64}$Ni+$^{208}$Pb and hot fusion $^{48}$Ca+$^{232}$Th reactions.   Table \ref{ComHotCold} presents comparison of the  measured ER cross sections $\sigma_{\rm ER}$,  excitation energy $E^*_{\rm CN}$  corresponding to the measured data and fission barrier $B_f$  of the CN calculated in Ref. \cite{Kowal2010,Jachimowicz2021},  as well as fusion probability $P_{\rm CN}$ for   the cold fusion  $^{64}$Ni+$^{208}$Pb  (from \cite{Giardina2000}) and hot fusion  $^{48}$Ca+$^{232}$Th  (this work) reactions.
\begin{figure}[ht]
\centering
\includegraphics[width=1.0\linewidth]{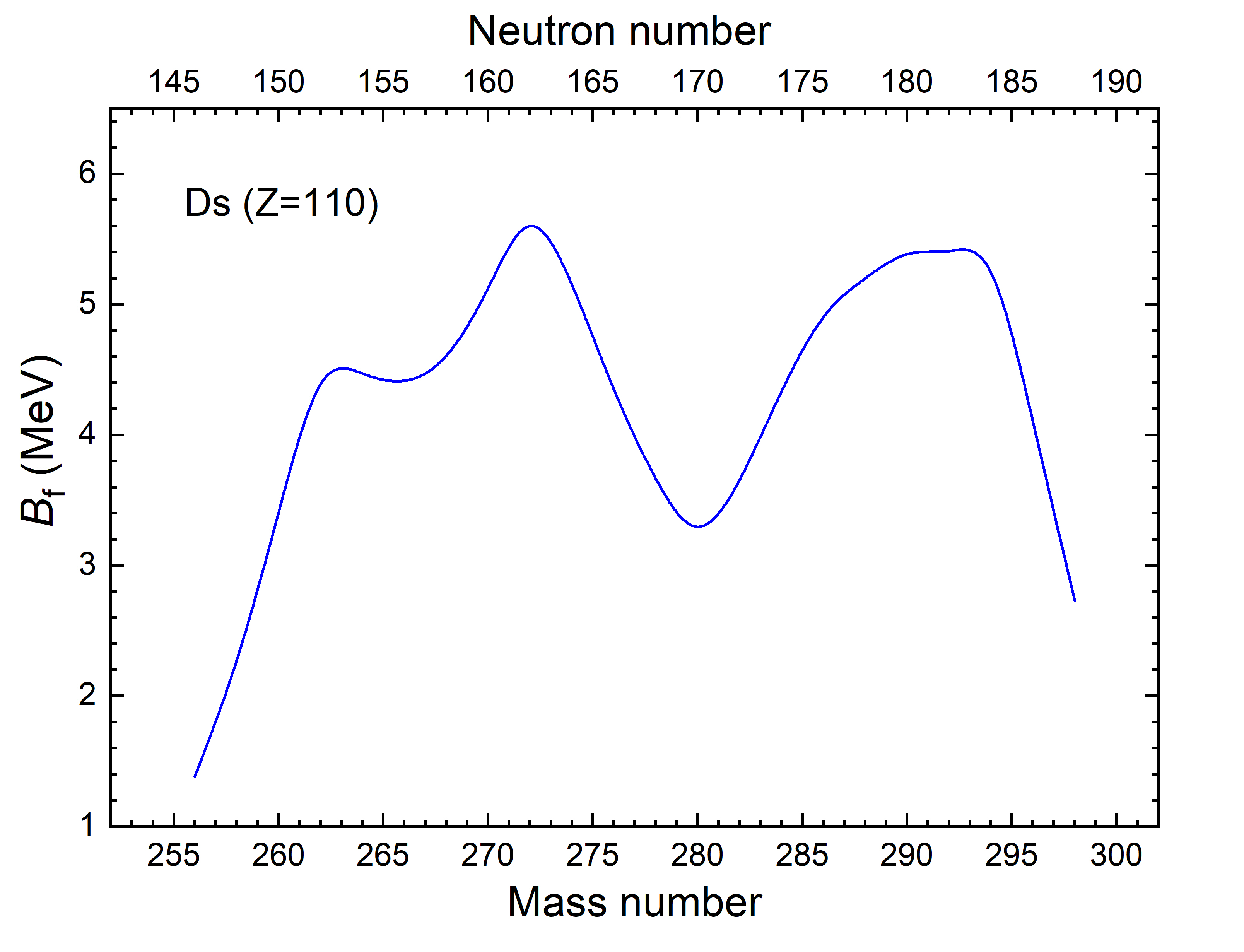}
\vspace{-0.75 cm}
\caption{(Color online) Fission barrier $B_f$ for darmstadtium ($Z$=110) obtained from \cite{Kowal2010} as a function of the mass and neutron numbers.}
\label{FisBar110}
\end{figure}

The main strong effect causing the large difference in the measured ER cross sections for the above mentioned reactions is a difference in the fission barrier providing the stability against fission. The fission barrier of the Ds isotopes is very sensitive to the neutron number due to the shell effects of its proton and neutron subsystems. Fig. \ref{FisBar110} shows the dependence of the  fission barrier $B_f$ on the mass and neutron numbers for the Ds isotopes. The isotopes $^{272}$Ds and $^{280}$Ds are formed as a CN in the  $^{64}$Ni+$^{208}$Pb and $^{48}$Ca+$^{232}$Th reactions, respectively. The neutron number of $^{272}$Ds is equal to magic number 162  for neutrons, consequently, the fission barrier has a maximum value. The small value of the excitation $E^*_{\rm CN}$=12.7 MeV is favorable for the survival probability $W_{\rm sur}$ and the ER cross section is large $\sigma_{\rm ER}=15$ pb in spite of the small fusion probability $10^{-5}$ \cite{Giardina2000}.  
  
The neutron number ($N=170$) of $^{280}$Ds is between two magic numbers 162 and 184 for the neutron subsystem.  Therefore, the fission barrier has a minimum value (see Fig.  \ref{FisBar110}). As a result the survival probability $W_{\rm sur}$ is very small for the excitation energies $E^*_{\rm CN}$=40.37 MeV. 
   
The dependence of the fission barrier $B_f$ on the excitation energy $E^*_{\rm CN}$ is taken into account at calculations of the survival probability $W_{\rm sur}$:
\begin{equation}
    B_{f}=B_{\rm LD}-f\delta{W}
\end{equation}
where $B_{\rm LD}$ and $\delta{W}$ are the liquid-drop fission barrier and the effective shell-correction energy, respectively. The liquid-drop fission barrier is estimated by using Lublin-Strasbourg Drop model \cite{Lublin}. The ground-state shell correction energies and the parametrizations for liquid-drop fission barrier are using the mass table of M\"{o}ller et al. \cite{MolNix1995}. The dependence of the  fission barrier on the excitation energy $E^*_{\rm CN}$ and angular momentum $\ell$ of the CN can be taken into account as in Ref.~\cite{Giardina2018}, where the correction factor $f$ was written as $h(T) q(\ell)$,
\begin{eqnarray}\label{hT}
    h(T)=\{1+\exp[(T-T_0)/d]\}^{-1}
\end{eqnarray}
and
\begin{eqnarray}\label{qL}
    q(\ell)=\{1+\exp[(\ell-\ell_{1/2})/\Delta\ell]\}^{-1}.
\end{eqnarray}
In Eq.(\ref{hT}) $T=\sqrt{E^*_{CN}/a}$ is nuclear temperature, $d=0.3$ MeV is the rate of washing out the shell corrections with the temperature, $T_0=1.16$ MeV is the value at which the damping factor $h(T)$ is reduced by $1/2$; in Eq.(\ref{qL}), $\Delta\ell=3\hbar$ is the rate of washing out the shell corrections with the angular momentum, $\ell_{1/2}=20\hbar$ is the value at which the damping factor $q(\ell)$ is reduced by $1/2$. To calculate the level density parameter $a$, Ignatyuk’s prescription \cite{Ignatyuk1975} was used.
The decrease of the neutron numbers  due to evaporation neutrons from $^{280}$Ds in competition with the fission process leads to   increase of the fission barrier. Therefore, a formation of  the isotope $^{176}$Ds formed in the  4n de-excitation channel has relatively large cross section in comparison with the 2n and 3n channels. The measured ER cross section of the 5n channel is lower than 4n channel since the fission barrier decreases by the increase of the excitation energy $E^*_{\rm CN}$.

\section{Conclusions}

The ER cross section of at the synthesis of $^{276}$Ds by the 3n, 4n and 5n 
de-excitation  channels in the $^{48}$Ca+$^{232}$Th reactions has been calculated and the results of this work is compared with the recent experimental from Ref. \cite{Oganessian2023,Oganessian2024} obtained in the SHE factory at 
JINR (Dubna, Russia). 
The maximal value 0.7 pb of the ER cross section  was observed for the 4n channel. In spite of the large probability of the CN formation in this reaction, the ER cross section is much  smaller (20 times) than the experimental data obtained in the cold fusion reactions $^{64}$Ni+$^{208}$Pb for the 1n channel by S. Hofmann {\it et al.} \cite{Hofmann1995}. This fact is explained by the dependence of the fission barrier $B_f$ on the mass number $A$ for the given element Ds ($Z=110$). The isotopes $^{272}$Ds and $^{280}$Ds are formed as a CN in the  $^{64}$Ni+$^{208}$Pb and  $^{48}$Ca+$^{232}$Th reactions, respectively. The neutron number of $^{272}$Ds is equal to magic number 162  for neutrons, consequently, the fission barrier has a maximum value. The small value of the excitation $E^*_{\rm CN}$=12.7 MeV is favorable for the survival probability $W_{\rm sur}$ and the ER cross section is large $\sigma_{\rm ER}=15$ pb in spite of the small fusion probability $10^{-5}$ \cite{Giardina2000}. The neutron number ($N=170$) of $^{280}$Ds is between two magic numbers 162 and 184 for the neutron subsystem. The decrease of the fission barrier $B_f$ at large excitation energies causes smallness of the ER cross section of $x$n channels with $x>4$ in hot fusion reactions.

\bibliography{references}

\end{document}